# Parallel Programming Model for the Epiphany Many-Core Coprocessor Using Threaded MPI

James A. Ross
Engility Corporation
Aberdeen Proving Ground, MD
james.ross@engilitycorp.com

David A. Richie
Brown Deer Technology
Forest Hill, MD
drichie@browndeertechnology.com

Song J. Park
U.S. Army Research Laboratory
Aberdeen Proving Ground, MD
song.j.park.civ@mail.mil

Dale R. Shires
U.S. Army Research Laboratory
Aberdeen Proving Ground, MD
dale.r.shires.civ@mail.mil

## ABSTRACT
The Adapteva Epiphany many-core architecture comprises a 2D tiled mesh Network-on-Chip (NoC) of low-power RISC cores with minimal uncore functionality. It offers high computational energy efficiency for both integer and floating point calculations as well as parallel scalability. Yet despite the interesting architectural features, a compelling programming model has not been presented to date. This paper demonstrates an efficient parallel programming model for the Epiphany architecture based on the Message Passing Interface (MPI) standard. Using MPI exploits the similarities between the Epiphany architecture and a conventional parallel distributed cluster of serial cores. Our approach enables MPI codes to execute on the RISC array processor with little modification and achieve high performance. We report benchmark results for the threaded MPI implementation of four algorithms (dense matrix-matrix multiplication, N-body particle interaction, a five-point 2D stencil update, and 2D FFT) and highlight the importance of fast inter-core communication for the architecture.

## Categories and Subject Descriptors
C.1.4 [**Parallel Architectures**]: Miscellaneous;
D.1.3 [**Concurrent Programming**]: Parallel Programming

## 1. INTRODUCTION
The emergence of a wide range of parallel processor architectures continues to present the challenge of identifying an effective programming model that provides access to the capabilities of the architecture while simultaneously providing the programmer with familiar, if not standardized, semantics and syntax. The programmer is frequently left with the choice of using a non-standard programming model specific to the architecture or a standardized programming model that yields poor control and performance.

The Adapteva Epiphany MIMD architecture is a scalable 2D array of RISC cores with minimal un-core functionality connected with a fast 2D mesh Network-on-Chip (NoC). Processors based on this architecture exhibit good energy efficiency and scalability via the 2D mesh network, but require a suitable programming model to fully exploit the architecture. The 16-core Epiphany III coprocessor has been integrated into the Parallella minicomputer platform where the RISC array is supported by a dual-core ARM CPU and asymmetric shared-memory access to off-chip global memory. Figure 1 shows the high-level architectural features of the coprocessor. Each of the 16 Epiphany III mesh nodes contains 32 KB of shared local memory (used for both program instructions and data), a mesh network interface, a dual-channel DMA engine, and a RISC CPU core. Each RISC CPU core contains a 64-word register file, sequencer, interrupt handler, arithmetic logic unit, and a floating point unit. Each processor tile is very small at 0.5 mm$^2$ on the 65 nm process and 0.128 mm$^2$ on the 28 nm process. Peak single-precision performance for the Epiphany III is 19.2 GFLOPS with a 600 MHz clock. Fabricated on the 65 nm process, the Epiphany III consumes 594 mW for an energy efficiency of 32.3 GFLOPS per watt [Olofsson, personal communication]. The 64-core Epiphany IV, fabricated on the 28 nm process, has demonstrated energy efficiency exceeding 50 GFLOPS per watt [14].

The raw performance of currently available Epiphany coprocessors is relatively low compared to modern high-performance CPUs and GPUs; however, the Epiphany architecture provides greater energy efficiency and is designed to be highly scalable. The published architecture road map specifies a scale-out of the architecture to exceed 1,000 cores in the near future and, shortly thereafter, tens of thousands of cores with an energy efficiency approaching one TFLOPS per watt. Within this context of a highly scalable architecture with high energy efficiency, we view it as a competitive processor technology comparable to GPUs and other coprocessors.

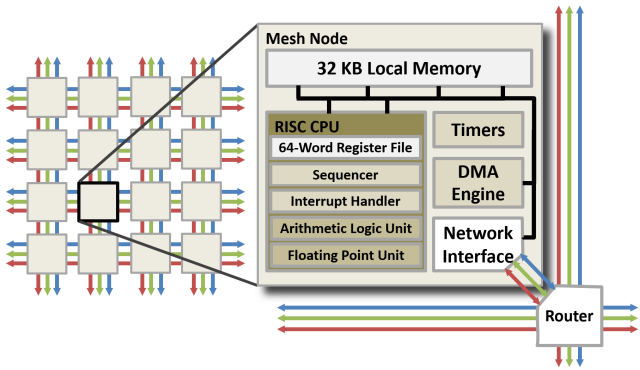

**Figure 1. Adapteva Epiphany III architecture diagram**

While architectural energy efficiency is important, achievable performance with a compelling programming model is equally, if not more, important. Key to performance with the Epiphany architecture is data re-use, requiring precise control of inter-core communication since the architecture does not provide a hardware cache at any level. The cores can access off-chip mapped memory with a significant performance penalty in both latency and bandwidth relative to accessing neighboring core memory.

When developing parallel applications, the parallel programming model and API must match the architecture lest it becomes overly complicated or unable to achieve good performance. By leveraging the standard MPI programming model, Epiphany software developers inherit a multi-decadal historical legacy of refined software design and domain decomposition considerations. For the Epiphany architecture, an MPI programming model is a better choice than OpenCL, which primarily targets GPUs. MPI is also superior to OpenMP since shared memory exhibits significant NUMA issues. Neither OpenCL nor OpenMP provide a mechanism for controlling inter-core data movement, which is critical to achieving high performance for anything but trivially parallel applications on this processor. Other programming models such as partitioned global address space (PGAS) using Unified Parallel C (UPC) may have merit with the Epiphany architecture; however, it requires an extension to ANSI C and removes some explicit control from the programmer.

In this paper, we demonstrate that threaded MPI exhibits the highest performance reported using a standard parallel API on the Epiphany architecture. Threaded MPI allows algorithm design to closely follow the methods for distributed parallel processors, and in some cases the code can be re-used with minimal modifications. Our previous demonstration of threaded MPI SGEMM on the Epiphany architecture [16] has been improved and the work has been extended to four algorithms used for benchmarking: SGEMM, iterative N-body particle interaction, an iterative 5-point 2D stencil update, and a 2D fast Fourier transform (FFT). Benchmarks for on-chip performance demonstrate that threaded MPI can be broadly applied to different algorithms. The peak performance achieved for each benchmark compares favorably with results reported for related benchmarks on other coprocessors, including the Intel Xeon Phi and Teraflops Research Chip.

## 2. THREADED MPI

Threaded MPI was developed to provide an extremely lightweight implementation of MPI appropriate for threads executing within the restricted context of the Epiphany RISC cores. Until now, the application developer for the Epiphany architecture was left with an option to either use a standard API, e.g., OpenCL, for trivially parallel applications or use nonstandard syntax and semantics to achieve high performance. The use of threaded MPI enables the use of a familiar and standard parallel programming API to achieve high performance and platform efficiency. The MPI programming model is well suited to the Epiphany architecture. The use of MPI for programming the Epiphany architecture was first suggested with a simple proof-of-concept demonstration in 2013 [12], and it is somewhat surprising that this line of research is only now being more thoroughly explored on this platform.

Here, threaded MPI is distinguished from conventional MPI implementations by two critical differences, driven by the fact the device must be accessed as a coprocessor and each core executes threads within a highly constrained set of resources. As a result, the cores are not capable of supporting a full process image or program in the conventional sense, and therefore the conventional MPI model of associating MPI processes to concurrently executing programs is not possible. Instead, coprocessor offload semantics must be used to launch concurrent threads that will then employ conventional MPI semantics for inter-thread communication.

The practical consequence is that rather than launching an MPI job from the command line, a host program executing on the platform CPU initiates the parallel MPI code using a functional call; the `mpiexec` command is replaced with an analogous function call, `coprthr_mpiexec(int device, int np, void* args, size_t args_sz, int flags)`. This has the advantage of localizing the parallelism to a fork-join model within a larger application that may otherwise execute on the platform CPU, and multiple `coprthr_mpiexec` calls may be made from within the same application. From the host application executing on the platform CPU, explicit device control and distributed memory management tasks must be used to coordinate execution with the Epiphany coprocessor at a higher level. These host routines are separate from the MPI programming model used to program the coprocessor itself. The only practical consequence and distinction with MPI code written for Epiphany, compared with a conventional distributed cluster, is that the `main()` routine of the MPI code must be transformed into a thread and employ Pthread semantics for passing in arguments. Beyond this, no change in MPI syntax or semantics is required.

Threaded MPI exhibits the highest performance reported using a standard parallel API for the Epiphany architecture. A minimal subset of the MPI standard, shown in Table 1, was used in four applications with various computation and communication patterns as a demonstration of the flexibility of the programming model. The `MPI_Sendrecv_replace` call, which provides a mechanism to exchange data between cores, showed remarkable versatility. The call cannot be implemented with a zero-copy design, but rather requires a buffered transaction. The available memory is highly constrained on each Epiphany core, making a large internal MPI buffer extremely costly. Therefore, the implementation must transparently segment large transfers into many small DMA transactions while maintaining high overall performance.

**Table 1. List of MPI calls used in four applications**

| MPI Library Calls | |
|---|---|
| MPI_Init | MPI_Comm_free |
| MPI_Finalize | MPI_Cart_coords |
| MPI_Comm_size | MPI_Cart_shift |
| MPI_Comm_rank | MPI_Sendrecv_replace |
| MPI_Cart_create | |

## 3. APPLICATIONS

Our previous demonstration of threaded MPI SGEMM [16] has been extended to include four algorithms commonly used for benchmarking: SGEMM, N-body particle interactions, five-point 2D stencil update, and 2D FFT. Each of these applications exhibits different computational and communication scaling, as shown in Table 2. Scaling is shown with respect to either the side-length $n$ of the problem size or in the case of N-body particle interaction, the total size $N$ for the problem. The performance numbers reported in this paper are for on-chip time only, excluding the startup cost to initialize the cores and copy data from off-chip memory.

**Table 2. Computation and communication algorithm scaling**

| Application | Computation | Communication |
|---|---|---|
| Matrix multiplication | $O(n^3)$ | $O(n^2)$ |
| N-body | $O(N^2)$ | $O(N)$ |
| 5-point 2D stencil | $O(n)$ | $O(n)$ |
| 2D FFT | $O(n^2 \cdot \log_2(n))$ | $O(n^2)$ |

The internal MPI buffer size used for buffered transfer protocols like that employed by the MPI_Sendrecv_replace routine is tunable and was selected to achieve the highest performance for each application within the limitations of the 32 KB local memory per core. The specific buffer size selected was 1.5 KB used for matrix–matrix multiplication, 1 KB for N-body simulation, 256 bytes for the 5-point stencil calculation, and 512 bytes for the 2D FFT.

### 3.1 MPI Latency and Bandwidth

The MPI_Sendrecv_replace routine was used for inter-core communication in each of the four applications. The routine was critical to achieving high communication performance. Figure 2 shows the effective inter-core bandwidth of the routine for a 2D periodic network topology where every core transfers data to the west and receives data from the east. This regular communication pattern achieves the highest total on-chip bandwidth, whereas a random communication pattern would be limited by the bisection bandwidth. The performance is a function of the total amount of data transferred and the internal MPI buffer size. As expected, the effective bandwidth improves with increasing internal MPI buffer size, with peak inter-core bandwidth performance for typical application transfers approaching 1,000 MB/s, which is equivalent to 80% of the theoretical peak of about 1,250 MB/s for a single-channel DMA transfer for the Epiphany III.

These data were fit to a standard alpha–beta model for MPI communication modified to account for the latency dependence on the internal MPI buffer size. Specifically, the latency will increase with the number of internal DMA transactions used to communicate a message of a given size. As the internal MPI buffer size is increased, the number of required transactions will decrease proportionately. Taking this into account, the total communication time $T$ is modeled as $T = \alpha_0 + \alpha_1 \cdot k + \beta \cdot m$, where $\alpha_0$ is the fixed latency of the MPI call, $\alpha_1$ is the latency per buffered DMA transaction, and $\beta^{-1}$ is the bandwidth. Here, $m$ is the message size in bytes and $k$ is the number of internal DMA transactions, calculated as $m$ divided by the MPI buffer size. The measured data fit to determine the values for the latency and bandwidth parameters and found to be $\alpha_0 = 1{,}216$ ns, $\alpha_1 = 309$ ns, $\beta^{-1} = 1{,}250$ MB/s.

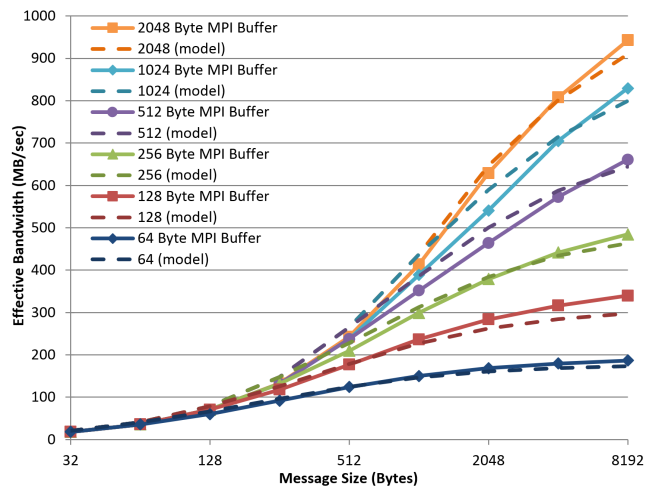

**Figure 2: Effective inter-core bandwidth of the MPI_Sendrecv_replace call for different MPI buffer sizes**

Significantly, the Epiphany architecture bandwidth scales with the number of on-chip cores. This differs from many other CPU architectures and the Xeon Phi where the effective inter-core bandwidth for an MPI code is limited by the bandwidth to main memory and may have higher clock latency due to the complex cache hierarchy.

Future improvements in the threaded MPI implementation could include a crossover point that replaces the use of the DMA engine in favor of direct inter-core write operations for small message sizes. This would have greatest effect on performance of grid applications where small messages of only boundary data are shared between neighboring cores.

### 3.2 Matrix–Matrix Multiplication

The matrix–matrix multiplication MPI code based on Cannon's algorithm for square matrices was previously developed for a conventional parallel cluster [9] and adapted here with minimal modification. In the context of the Epiphany coprocessor, the first communication step—which skewed the submatrices data across the cores—was removed as it is unnecessary when using a temporary copy of the host matrices. Instead, the submatrices were read in from main memory preskewed and the B submatrix was transposed for a more efficient memory access pattern within the inner loop of the matrix–matrix multiply. The final communication step to reorder the data was also removed since it was unnecessary because it was an intermediate copy of the host data. The three inner loops of the matrix–matrix multiplication were unrolled by four and a fused multiply–add (FMA) GCC built-in function was used to force the compiler to generate the most efficient code. The inner loop then demonstrated operation at the peak performance of the core. This by itself is significant—that the compiler and software stack can generate high-performance programs from C code.

The reported benchmarks used an internal MPI buffer size of 1.5 KB, although little performance improvement is seen beyond 512 bytes. The performance of the matrix–matrix multiplication code in GFLOPS is calculated using the formula $2 \cdot n^3/t$, where $n$, is the matrix size in a single dimension and $t$ is the total execution time in nanoseconds. Figure 3 shows the overall measured performance of the full matrix–matrix calculation on 16 cores including the blocking communication time between the cores. For each workload, the *relative time* spent for computation and communication is estimated and shown visually as a relative fraction of the represented performance. This algorithm exhibits a roughly even breakdown between computation and communication costs, with slightly more time spent on computation at larger workloads. Despite the communication taking a significant fraction of the execution time, the application is able to achieve 12.02 GFLOPS. This is about 63% of the peak performance of the processor and 74% (12.02 GFLOPS vs. 16.2 GFLOPS) of the performance achieved in [23] for a 4×4 array of cores on the Epiphany IV processor, which used low-level optimized assembly and a double-buffer scheme. This performance result exceeds 20 GFLOPS per watt.

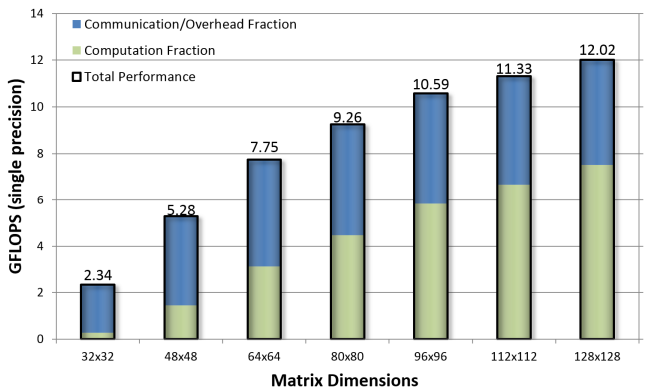

Figure 3. On-chip matrix–matrix multiplication performance

## 3.3 N-body Particle Interaction

The N-body code was based on an existing MPI implementation of the algorithm [13]. The code was extended from two to three dimensions; the `MPI_Isend` and `MPI_Irecv` pair was replaced with `MPI_Sendrecv_replace`; and then the code was optimized for performance. This algorithm has the most favorable compute-to-communication ratio of the four algorithms examined in this work; the application was generally not communication-bound and can be considered a best-case scenario for nontrivial parallelism. It is also iterative so that once the particle data is on-chip, there is a significant amount of computation. Unlike the other applications, the N-body calculation uses a 1D communication topology so that data moved in a single scan line cycle. A network topology based on a continuous fractal space-filling curve was explored but had a negligible effect on the overall performance. High performance was achieved by unrolling the outer compute loop by eight, explicitly using the FMA built-in function, and using a fast approximation for the inverse square root. For each particle position update, the cores shift the working set of particle positions and masses to the neighboring core. Once the working set has completely cycled through all the cores, the working set is updated with the new particle positions and the next time step iteration takes place.

The performance in GFLOPS is calculated using the convention $20 \cdot i \cdot N^2/t$ where $i$ is the number of time step iterations, $N$ is the number of particles, and $t$ is the total execution time in nanoseconds. The convention considers the inverse square root to be two floating point operations although it takes multiple cycles to compute. Benchmarks reported used an internal MPI buffer of 1,024 bytes. As shown in Figure 4, the *relative time* spent on communication was significantly low so that the internal buffer could have been much smaller. Any internal MPI buffer of more than 64 bytes would have been sufficient to achieve negligible relative communication time beyond 1,024 particles. The highest measured performance of 8.28 GFLOPS is 43% of theoretical peak. The use of a software square root operation and deviation from a one-to-one ratio of multiply/add operations make it impossible to achieve the theoretical peak for this particular algorithm.

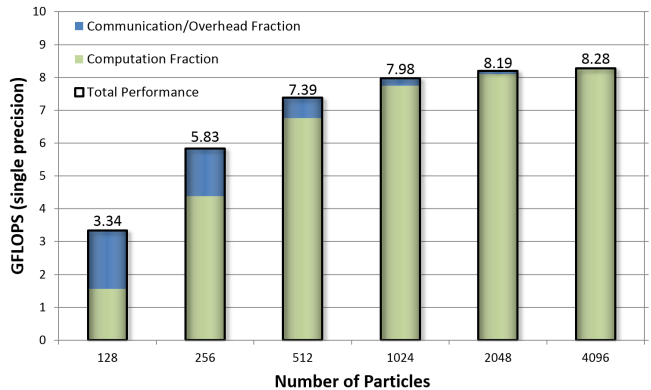

Figure 4. On-chip N-body particle interaction performance

The on-chip work size is limited to fewer than 5,000 particles. Larger calculations can be performed by streaming data from global memory and throughput only marginally decreases (less than 5%).

## 3.4 Five-Point 2D Stencil

The 5-point 2D stencil update was implemented using a conventional parallelization with MPI. Other researchers have used this grid application to evaluate performance on the Epiphany [23] and similar architectures [22]. This analog for many finite difference approximations combines the grid point itself along with neighboring data in the cardinal directions, scales these values by a constant, and stores the result for subsequent iterations. Of all of the applications, this algorithm has the least favorable computation-to-communication ratio so that performance is mostly limited by communication. However, the algorithm is iterative so there is a significant amount of computation once the data is on-chip. For each point, there is one multiplication and four FMA operations using five values loaded from memory. Implementation of the benchmark used conventional parallelization with MPI on a 2D network topology. Communication occurs among each of the four cardinal directions per iteration. The 2D computational domain is distributed across all cores such that it mirrors the physical network layout. Network domain and physical domain boundaries are not shared and the data values are kept fixed. Each communication buffer is copied into a temporary storage buffer before sending and receiving. The inner two loops of the algorithm, iterating by rows and then by columns, were optimized by unrolling each by four so that a 4×4 block of data and edges were loaded into registers for optimal data re-use. Additionally, after each column block shift right, the previous edge values were reused for the next block calculation. The performance in GFLOPS is calculated using the formula

$9 \cdot i \cdot n^2/t$ where $i$ is the number of iterations, $n$ is the problem size along one dimension, and $t$ is the execution time in nanoseconds.

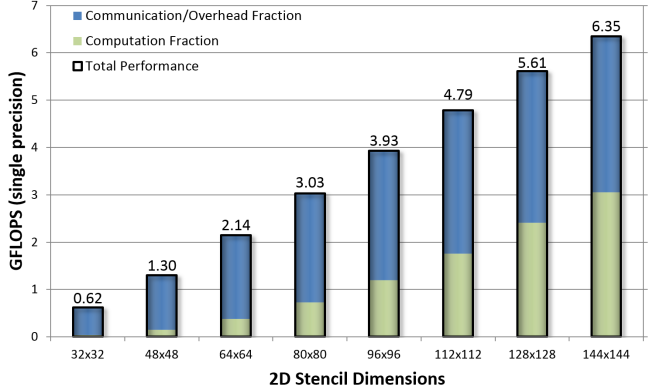

Figure 5. On-chip 5-point 2D stencil performance

The reported benchmarks used an internal MPI buffer of 256 bytes. For each workload, the *relative time* spent for computation and communication is estimated and shown visually as a relative fraction of the represented performance in Figure 5. The highest measured performance of 6.35 GFLOPS is 33% of theoretical peak. Part of the reason for the low performance is that the effective inter-core bandwidth is low due to the small buffer sizes. In the case of the 128×128 workload, each edge buffer is 128 bytes. As shown in Figure 2, one should not expect very high communication performance but rather less than 100 MB/s.

## 3.5 2D Fast Fourier Transform

The complex 2D FFT application is parallelized over stripes and performs a 1D radix-2 decimation-in-time (DIT) FFT, a corner turn, an additional radix-2 DIT FFT, and a final corner turn using the Cooley–Tukey algorithm. The implementation of the benchmark used a conventional parallelization with MPI. At small workloads, the performance is dominated by the corner turn communication. For improved performance the inner loop of the in-place radix-2 DIT code was unrolled by two. This unroll is unlike the other applications and nontrivial. Additionally, the complex data type is less amenable to optimization with FMA built-in functions. The performance in GFLOPS is calculated using the standard formula for complex 2D FFT used by the FFTW group of $5 \cdot n^2 \cdot \log_2(n^2)/t$ where $n$ is the problem size along one dimension and $t$ is the execution time in nanoseconds. (Note that this is not a precise FLOP count but, rather, a convenient scaling for comparison to other algorithms.)

The reported benchmarks used an internal MPI buffer of 512 bytes. For each workload, the *relative time* spent for computation and communication is estimated and shown visually as a relative fraction of the represented performance in Figure 1. The highest measured performance of 2.50 GFLOPS is 13% of theoretical peak. Despite exhibiting the lowest percent of theoretical peak of the four algorithms examined, the result is quite good in comparison to other CPU architectures, particularly on using a power-efficiency metric.

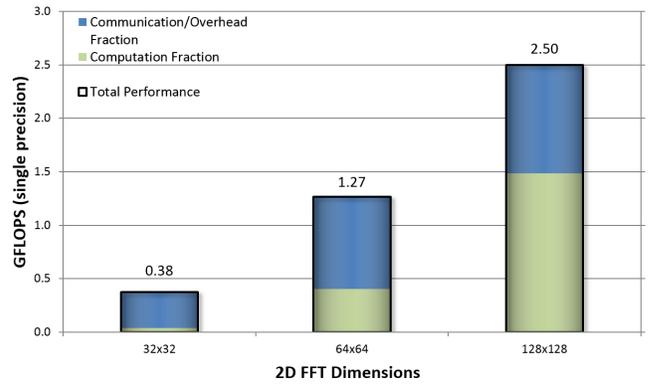

Figure 6. On-chip 2D fast Fourier transform performance

## 4. DISCUSSION

The benchmarks for on-chip performance demonstrate that threaded MPI is generally applicable for a number of applications on the Epiphany architecture. The benchmark trends suggest that the performance would improve if the Epiphany cores had more than 32 KB of available shared local memory, and the Epiphany road map indicates that this change will occur in future generations. Typically, the program and stack use approximately 16 KB of the available memory, so doubling the shared local memory to 64 KB would effectively triple the usable application memory to about 48 KB. This generally has the effect of increasing the amount of local work per core, reducing the relative amount of communication, and improving overall performance and performance per core. Alternatively, larger programs could be executed with more core memory.

The lack of a hardware cache at any level on the Epiphany architecture has many benefits, but at the expense of requiring good software design. Direct access to the local shared memory is very fast, roughly equivalent to the speed of L1 cache. Removing a cache allows for a higher core count within the same die area or lower power consumption. These design decisions directly impact the ability to scale to thousands of cores in the near future.

The demonstrated performance on Epiphany, as a percentage of peak performance, compares very favorably with other coprocessors such as the Intel Xeon Phi and Intel Teraflops Research Chip. The Intel Xeon Phi achieves 85% of peak performance on SGEMM using the Intel Math Kernel Library (MKL), although achievable performance using a standard parallel API such as OpenMP or MPI is typically much lower. Vangal et al. [22] reported the 80-core Intel Teraflops Research Chip achieved 37.5% of peak performance on SGEMM, 2.73% of peak performance on 2D FFT using the same algorithm, and 73.3% of peak on a stencil code with a different algorithm and communication pattern.

## 5. RELATED WORK

Adapteva's Epiphany SDK includes ANSI C/C++ support [26]. Other methods of programming include COPRTHR SDK [15], Erlang [4], Array Manipulation Language [20], and CAL Actor Language with Network Language [5]. A model based on message passing paradigm is not known to have been implemented.

In terms of applications, a few studies have surveyed programming and performance of the Epiphany architecture. Matrix multiplication was tested in Sapir's work [17]. The 2D FFT was benchmarked in Sapir's other work [18]. The back-projection algorithm was implemented on Epiphany in Zain-Ul-Abdin et al. [25]. Stencil and matrix multiplication was analyzed for the 64-core Epiphany in Varghese et al. [23]. Hash decryption on Parallella board was tested in Malvoni and Knezovic's work [7]. However, these previous studies used the Epiphany SDK directly and without a standard programming model and API.

As for hardware, several prior works exist in tiled many-core architectures. Intel has developed tile-based, many-core prototypes Teraflop Research Chip [22] and Single-chip Cloud Computer (SCC) [6]. Tilera has produced a tiled cache-coherent TILE64 [24] inspired by MIT's RAW processor [19]. Ambric introduced RISC with a massively parallel processor array chip Am2045 [1]. Scorpio [2] is an academic 36-core chip design at the Massachusetts Institute of Technology. STMicroelectronics has designed a many-core processor Platform 2012 [11], also known as STHORM [8]. Kalray has released a high-performance, low-power MPPA-256 chip [3]. Typically, these architectures have low-level primitives for inter-core communication, some basic communication libraries, and some MPI implementations at various feature levels.

Message passing API RCCE [10] and lightweight MPI implementation RCKMPI [21] were developed for the SCC processor. Similar to our work, the many-core processor was programmed with MPI codes. Mattson et al. [10] argue that the SCC processor is a message passing chip at its core and thereby the most efficient programming model should have the ability to send messages between cores. However, there is a notable distinction of the Epiphany chip being represented as a coprocessor in currently available platforms.

## 6. CONCLUSIONS
We demonstrated, using a standard API, that threaded MPI provides a programming model capable of achieving high performance and efficiency for a range of parallel applications on the Epiphany many-core architecture without employing hand-tuned assembly or custom inter-core communication code. The proposition of viewing the architecture as a device-scale parallel networked cluster with a 2D communication topology was further validated. We demonstrated that previously written MPI software can be ported to the Epiphany with little modification, simplifying the processes of domain decomposition and application development. Benchmark results using a conventional MPI implementation compare favorably with the matrix–matrix multiplication, 2D FFT, and 2D stencil update performance on similar architectures.

## 7. FUTURE WORK
Future work will investigate the application of non-blocking zero-copy MPI communication calls to enable an overlap of computation and communication. This is expected to increase performance for the matrix–matrix multiplication and stencil applications. A more complete network performance model for MPI, only briefly discussed in Section 3.1, will be developed and compared to other modern architectures such as the Intel Xeon Phi. Additional applications will be explored.

## 8. ACKNOWLEDGMENTS
The authors wish to acknowledge the U.S. Army Research Laboratory-hosted Department of Defense Supercomputing Resource Center for its support of this work.

## 9. REFERENCES

[1] Butts, M., Jones, A.M. and Wasson, P. 2007. A structural object programming model, architecture, chip and tools for reconfigurable computing. *Field-Programmable Custom Computing Machines (FCCM)* (Apr. 2007), 55–64.

[2] Daya, B.K., Chen, C.-H.O., Subramanian, S., Kwon, W.-C., Park, S., Krishna, T., Holt, J., Chandrakasan, A.P. and Peh, L.-S. 2014. SCORPIO: A 36-core research chip demonstrating snoopy coherence on a scalable mesh NoC with in-network ordering. *Proceeding of the 41st Annual International Symposium on Computer Architecture (ISCA'14)* (2014), 25–36.

[3] Dinechin, B.D. de, Massas, P.G. de, Lager, G., Léger, C., Orgogozo, B., Reybert, J. and Strudel, T. 2013. A distributed run-time environment for the Kalray MPPA®-256 integrated manycore processor. *Procedia Computer Science*. 18, (2013), 1654–1663.

[4] Erlang-OTP and the Parallella Board: 2015. *https://www.parallella.org/2015/03/20/erlang-otp-and-the-parallella-board/*. Accessed: 2015-03-25.

[5] Gebrewahid, E., Yang, M., Cedersjo, G., Abdin, Z.U., Gaspes, V., Janneck, J.W. and Svensson, B. 2014. Realizing efficient execution of dataflow actors on manycores. *Proceedings of the 2014 12th IEEE International Conference on Embedded and Ubiquitous Computing (EUC '14)* (Aug. 2014), 321–328.

[6] Howard, J., Dighe, S., Hoskote, Y., Vangal, S., Finan, D., Ruhl, G., Jenkins, D., Wilson, H., Borkar, N., Schrom, G. and others 2010. A 48-core IA-32 message-passing processor with DVFS in 45nm CMOS. *International Solid-State Circuits Conference Digest of Technical Papers (ISSCC)* (2010), 108–109.

[7] Malvoni, K. and Knezovic, J. 2014. Are your passwords safe: energy-efficient bcrypt cracking with low-cost parallel hardware. *8th USENIX conference on Offensive Technologies (WOOT'14)* (2014).

[8] Marongiu, A., Capotondi, A., Tagliavini, G. and Benini, L. 2013. Improving the programmability of STHORM-based heterogeneous systems with offload-enabled OpenMP. *First International Workshop on Many-core Embedded Systems (MES'13)* (2013), 1–8.

[9] Matrix-Matrix Multiply: *https://svn.mcs.anl.gov/repos/performance/benchmarks/mpi_overlap/matmul.c*. Accessed: 2015-03-23.

[10] Mattson, T.G., Riepen, M., Lehnig, T., Brett, P., Haas, W., Kennedy, P., Howard, J., Vangal, S., Borkar, N., Ruhl, G. and others 2010. The 48-core SCC processor: the programmer's view. *ACM/IEEE International Conference for High Performance Computing, Networking, Storage and Analysis (SC '10)* (2010), 1–11.

[11] Melpignano, D., Benini, L., Flamand, E., Jego, B., Lepley, T., Haugou, G., Clermidy, F. and Dutoit, D. 2012. Platform 2012, a many-core computing accelerator for embedded SoCs: performance evaluation of visual analytics applications. *49th Annual Design Automation Conference (DAC '12)* (2012), 1137–1142.

[12] MPI "lite" proof-of-concept: *https://parallella.org/forums/viewtopic.php?f=17&t=95*. Accessed: 2015-03-26.

[13] N-body program using pipelining algorithm: *http://www.mcs.anl.gov/research/projects/mpi/usingmpi/examples-usingmpi/advmsg/nbodypipe_c.html*. Accessed: 2015-03-23.

[14] Olofsson, A., Nordström, T. and Ul-Abdin, Z. 2014. Kickstarting high-performance energy-efficient manycore architectures with Epiphany. *arXiv preprint arXiv:1412.5538*. (2014).

[15] Richie, D. 2013. *COPRTHR API Reference*. Brown Deer Technology.

[16] Richie, D., Ross, J., Park, S. and Shires, D. 2015. Threaded MPI programming model for the Epiphany RISC array processor. *Accepted for publication in Special Issue of the Elvesier Journal of Computational Science* (2015).

[17] Sapir, Y. 2012. *Scalable parallel multiplication of big matrices*. Adapteva, Inc.

[18] Sapir, Y. 2012. *Using a scalable parallel 2D FFT for image enhancement*. Adapteva, Inc.

[19] Taylor, M.B., Lee, W., Miller, J., Wentzlaff, D., Bratt, I., Greenwald, B., Hoffmann, H., Johnson, P., Kim, J., Psota, J. and others 2004. Evaluation of the RAW microprocessor: An exposed-wire-delay architecture for ILP and streams. *31st Annual International Symposium on Computer Architecture (ISCA '04)*. (2004), 2–13.

[20] The Apl to C compiler aplc is now ported to the Parallella: *http://forums.parallella.org/viewtopic.php?f=13&t=338#p2016*. Accessed: 2015-03-25.

[21] Ureña, I.A.C., Riepen, M. and Konow, M. 2011. RCKMPI–lightweight MPI implementation for Intel's Single-chip Cloud Computer (SCC). *18th European MPI Users' Group conference on Recent advances in the message passing interface (EuroMPI'11)* (2011), 208–217.

[22] Vangal, S.R. et al. 2008. An 80-tile sub-100-W teraFLOPS processor in 65-nm CMOS. *IEEE Journal of Solid-State Circuits*. 43, 1 (Jan. 2008), 29–41.

[23] Varghese, A., Edwards, B., Mitra, G. and Rendell, A.P. 2014. Programming the Adapteva Epiphany 64-Core network-on-chip coprocessor. *2014 IEEE International Parallel & Distributed Processing Symposium Workshops (IPDPSW '14)* (May 2014), 984–992.

[24] Wentzlaff, D., Griffin, P., Hoffmann, H., Bao, L., Edwards, B., Ramey, C., Mattina, M., Miao, C.-C., Brown III, J.F. and Agarwal, A. 2007. On-chip interconnection architecture of the tile processor. *IEEE micro*. 27, 5 (Sep. 2007), 15–31.

[25] Zain-Ul-Abdin, Ahlander, A. and Svensson, B. 2013. Energy-efficient synthetic-aperture radar processing on a manycore architecture. *2013 42nd International Conference on Parallel Processing (ICPP '13)* (Oct. 2013), 330–338.

[26] *Epiphany Architecture Reference*. Technical Report #Rev. 14.03.11. Adapteva.